# Concealing arbitrary objects remotely with multi-folded transformation optics

B. Zheng[1,2,3†], H. A. Madni[1,2,3†], R. Hao[2], X. Zhang[2], X. Liu[1], E. Li[2*] and H. Chen[1,2,3*]

[1]State Key Laboratory for Modern Optical Instrumentation, Zhejiang University, Hangzhou 310027, China

[2]College of Information Science and Electronics Engineering, Zhejiang University, Hangzhou 310027, China

[3]The Electromagnetics Academy at Zhejiang University, Zhejiang University, Hangzhou 310027, China

## Abstract

An invisibility cloak that can hide an arbitrary object external to the cloak itself has not been devised before. In this Letter, we introduce a novel way to design a remote cloaking device that makes any object located at a certain distance invisible. This is accomplished using multi-folded transformation optics to remotely generate a hidden region around the object that no field can penetrate and that does not disturb the far-field scattering electromagnetic field. As a result, any object in the hidden region can stay in position or move freely within that region and remain invisible. Our idea is further extended in order to design a remote illusion optics that can transform any arbitrary object into another one. Unlike other cloaking methods, this method would require no knowledge of the details of the object itself. The proposed multi-folded transformation optics will be crucial in the design of remote devices in a variety of contexts.

[†] Co-first author. These authors contributed equally to this work.

*Corresponding Author: liep@zju.edu.cn and hansomchen@zju.edu.cn.

The ingenious theory of transformation optics (TO) proposed by Pendry et al. [1] gives the ability to control electromagnetic waves. Due to this extraordinary feature, TO has many applications, such as cloaking devices [2-6], superlenses [7-8], antennas [9] etc. Among them , the most attractive is a cloaking device that can direct light to propagate smoothly around an object, making it disappear for the viewers [1-6]. Since the transformation-based cloak was firstly proposed, extensive theoretical and experimental works have been developed to demonstrate the functionality of this interesting device [10-18]. An important feature of this invisibility cloak is that the cloaking shell should enclose the object such that no incident wave can penetrate into the hidden region, and therefore it cannot work remotely.

In order to hide an object at a distance, Lai et al. [19] proposed the "anti-object" concept [20], a complementary "image" object that cancels the scattering field generated by a pre--specified object. Based on this method, a scheme for illusion optics [21] was further introduced to generate the illusion of transforming one object into another one. A key feature of a remote cloak is that it does not require the cloak shell to enclose the object. However, unlike Pendry's original cloak that can hide any arbitrary object, Lai's remote devices are designed for a particular object with known shape, size, and position. Therefore, the cloak can operate only for a specific object, and a little change in its position can destroy the exact restoration and cancellation of the optical path of the incident wave. Another approach to remote cloaking was investigated to design active exterior cloaks [22] that require advance knowledge of the phase information and of the incoming probing wave. In short, a remote cloaking device that can hide any arbitrary object is still elusive.

In this Letter, we propose a new recipe for a remote cloaking device that requires no knowledge of the hidden object's shape and material. Our designing technique is based on multi-folded transformation optics [23] to remotely produce a hidden region into which the incoming electromagnetic waves cannot penetrate, and that does not alter the far-field scattering field. Therefore, the device can still rendered invisible even with the existence of such hidden region. Most importantly, the proposed device is not based on the "anti-object" concept and the hidden object is not restricted to move within a certain range of the hidden region to avoid affecting the outer boundary field pattern. Our new designed device has no physical attachment with the hidden object so any object inside the hidden range can move freely. We further extend this idea to design an object-independent remote illusion optics that remotely creates the illusion of an object looking like another object. Full-wave finite element simulations validate the expected behavior of our proposed device.

Fig. 1(a) shows the schematics of a conventional cloak [1] that conceals an object inside the cloak region without altering the incoming field. This is achieved using a coordinate transformation to create a hole in the center such that no fields can penetrate inside the cloak but the outside field remains unaltered. Therefore, any object inside the cloak is invisible to the outside observers. A conventional remote cloaking [19], instead, cloaks a pre-specified object outside of the cloaking shell by using the "anti-object" concept. The idea is illustrated in Fig. 1(b), where an object with constitutive parameters $\varepsilon'(r)$ and $\mu'(r)$ is placed in free space (on the right), and a complementary image object with constitutive parameters $-\varepsilon'(r)$ and $-\mu'(r)$ is embedded in the complementary media (on the left). As a consequence, the object and the complementary media with anti-object cancel the scattering field generated by the object and the object becomes undetectable.

Unlike the methods mentioned above, in this paper we aim to create a hidden region at a distance, so that any object moving inside this hidden region can be invisible, as shown in Fig. 2(a). Therefore, this remote cloaking device is object-independent. Here, as an example, the remote cloaking device is made of two elements that create a hidden region, but the same method can be extended to an arbitrary number of elements. Fig. 2(b) depicts the effect of this object-independent remote cloaking device in free space. The figure shows the equivalent virtual space created by the device and how the wave can propagate through the cloak without any perturbations. Fig.2(c) illustrates the schematic diagram of our proposed remote cloaking device with two transformation steps. In the first step, a square shaped cloak filled with air placed in the virtual space $(x, y, z)$ is transformed into a conventional cloak (light blue segments). In the next step, the cloak is divided into four segments (delimited by dashed lines and denoted $A$, $B$, $C$ and $D$); the second transformation function on each segment is used to compress the whole segment as $A'$, $B'$, $C'$ and $D'$, respectively, so that the cloak is open and does not require enclosing the hidden object. As shown in Fig. 2(c), each segment is further divided into four regions. For example, region $A'$ contains the sub-region I (green area), represented by $(x_1, y_1, z_1)$, sub-region II (dark blue), denoted by $(x_2, y_2, z_2)$, sub-region III (orange), denoted by $(x_3, y_3, z_3)$, and sub-region IV (maroon) denoted by $(x_4, y_4, z_4)$. A zoomed-in view of the proposed system, clarifying the design methodology, is presented at the most right side of Fig. 2(c). Consequently, we compress the complete black dashed lined area into the green shaded sub-region (region I) and the dark blue shaded sub-region (region II). It should be noticed that the material parameters of region II contain the material parameters of region I and of the conventional square cloak. Thereafter, a folded transformation is used to compensate the discontinuity due to the compressing coordinate transformation. In this case, the sub-regions III and IV (complementary materials) are obtained with different parameters at different folding ranges. $x_n$, $y_n$ and $z_n$ indicate the coordinate system of each region, where $n = 1, 2, 3$ and $4$.

For the first step shown in Fig. 2(c), the constitutive parameters of the conventional cloak are obtained from ref. 6. The resulting permittivity and permeability tensors for the segment $A$ of the conventional square cloak can be expressed as:

$$\varepsilon_r = \mu_r = \begin{pmatrix} \dfrac{c}{a} & -\dfrac{b}{a} & 0 \\ -\dfrac{b}{a} & \dfrac{a^2 + b^2}{ac} & 0 \\ 0 & 0 & ac \end{pmatrix}, \qquad (1)$$

with

$$a = \frac{s_2}{s_2 - s_1}, b = \frac{y}{(x)^2} a s_1, c = \left(1 - \frac{s_1}{x}\right). \qquad (2)$$

The half of the side lengths of the inner and outer square are $s_1$ and $s_2$ respectively (see Fig. 2(c)). In the next step (second transformation in Fig. 2(c)), for the segment $A'$, the transformation equations of the sub-regions I and II in the Cartesian coordinates are given by:

$$\begin{aligned} x' &= x \\ y' &= \kappa y \\ z' &= z \end{aligned}, \qquad (3)$$

where $\kappa = \tan\beta/\tan\alpha$. The constitutive parameters for the sub-regions I and II are obtained from the Jacobian transformation matrix as follows:

$$\varepsilon_1' = \mu_1' = \begin{pmatrix} \dfrac{1}{\kappa} & 0 & 0 \\ 0 & \kappa & 0 \\ 0 & 0 & \dfrac{1}{\kappa} \end{pmatrix} \tag{4}$$

$$\begin{aligned} \varepsilon_2' &= \varepsilon_1' \cdot \varepsilon_r \\ \mu_2' &= \mu_1' \cdot \mu_r \end{aligned} \tag{5}$$

Moreover, the folded transformations are applied to the sub-regions III and IV. Taking the orange shaded area in the first quadrant as an example in sub-region III, leads to the transformation equation:

$$\begin{aligned} x_3' &= x \\ y_3' &= -y + 2\tau(P - x) \\ z_3' &= z \end{aligned}, \tag{6}$$

where $\tau = \tan\gamma = (\tan\alpha - \tan\beta)/2$ and $P$ is the intersection point of all the sub-regions at the horizontal-axis. The material parameters of the complementary material for this region are:

$$\varepsilon_3' = \mu_3' = \begin{pmatrix} -1 & 2\tau & 0 \\ 2\tau & -(4\tau^2 + 1) & 0 \\ 0 & 0 & -1 \end{pmatrix}. \tag{7}$$

Similarly, taking the maroon area in the second quadrant as an example in sub-region IV, gives the transformation equation and the constitutive parameters:

$$x' = x$$
$$y' = -y + 2\tau_1(x - P_1) \tag{8}$$
$$z' = z$$

$$\varepsilon_4' = \mu_4' = \begin{pmatrix} -1 & -2\tau_1 & 0 \\ -2\tau_1 & -(4\tau_1^2 + 1) & 0 \\ 0 & 0 & -1 \end{pmatrix} \tag{9}$$

where $\tau_1 = \tan\gamma_1 = (\tan\alpha_1 - \tan\beta_1)/2$ and $P_1$ is the coordinate value along the x-axis. The other parts of these complementary regions transform in a similar way, that for brevity we do not write in this Letter. In addition, due to the symmetry of the cloak, the permittivity and permeability tensors of the other device's domains can be obtained by rotating all of the corresponding tensors by π/2, π, and 3π/2.

The following section summarizes the full wave simulations of the proposed remote cloaking device by adopting the scattered TE mode with a frequency of 3 GHz and by using the 2D finite element simulator (COMSOL). For the closed square cloak, the simulation uses the following parameters: $s_1 = 0.075$ m and $s_2 = 0.1$ m, the compression ratio $\kappa = 0.8$ with $\alpha = 45°$, $\gamma = 42°$ and $\beta = 38.66°$ for the sub-regions I and II. Additionally, $\tau = 0.9$ with the same $\alpha$, $\gamma$ and $\beta$ as in the sub-region III and $\tau_1 = 2.7$ with $\alpha_1 = 71.57°$, $\gamma_1 = 69.68°$ and $\beta_1 = 67.38°$ for a different folding range of sub-region IV.

In Fig. 3, we consider the electric field components of the wave for objects of different shape and compare the field behaviors for each with and without the remote cloaking device. In Fig. 3(a), a PEC circle with a radius of 0.03 m is placed in the air and is subjected to a plane wave. In this case, the field is scattered. In Fig. 3(b), the remote cloaking device is used to cloak the PEC circle. It can be seen that the scattering field caused by the PEC circle (visible in Fig. 3(a)) is well minimized by the open cloaking device and the PEC circle becomes invisible. It should be noted that, although the remote cloaking device is open, the center field region it creates is a hidden region that no wave can penetrate, and therefore, any change of the objects does not affect the performance of the cloak. In order to verify this, we changed the hidden object to be a PEC square, and then a dielectric star shaped object with $\varepsilon_0 = 5$ and $\mu_0 = 1$. The results shown in Fig. 3(c-f) illustrate that the scattering caused by these different objects can be well minimized by the same remote object-independent cloak.

This multi-folded TO method can be similarly extended to design remote illusion optics. The real space of object-independent illusion optics is illustrated in Fig. 4(a). In this scenario, the remote device, embedded with the compressed object selected for the illusion, is used to create the hidden region. Fig. 4(b) demonstrates the illusion effect created by the illusion device. In general, any object (here a flower is shown as an example) inside the hidden region will turn into another object, e.g., a stick. The schematic diagram of the proposed illusion optics is shown in Fig. 4(c). This device is based on the remote cloaking device discussed earlier, with the addition of an illusionary object embedded inside the remote cloaking device. Here, as an example, we used a dielectric stick as embedded object. The stick is first transformed and adjusted into the segment $B$ of the conventional cloak, and further compressed in the second transformation step with the same equation as regions I and II for segment $B'$ of the proposed remote cloaking device.

Fig. 5(a-f) is the simulation result that demonstrates the functionality of the remote illusion device in transforming one image into another. In Fig. 5(a), a dielectric flower with $\varepsilon_f = 3$ and

$\mu_f = 1$ is placed at the center, with a plane wave incident from left to right. Fig. 5(b) shows the scattering pattern of the flower with the remote illusion device placed beside it, while Fig. 5(c) shows the scattering pattern of a dielectric stick of 0.02 m with $\varepsilon_s = 5$ and $\mu_s = 1$. Comparing the field patterns of Fig. 5 (a-c), it is evident that the remote illusion device makes the flower behave like if there were a stick. Furthermore, the device is omnidirectional. For example, Fig 5(d-f) shows the simulated results when the propagation direction of the incident plane wave is rotated by $\pi/4$. The material parameters and functionality in Fig 5(d), 5(e), and 5(f) are the same as in Fig. 5(a), 5(b) and 5(c), respectively. One can see that the scattering pattern of Fig. 5(e) is identical to Fig. 5(f), which is a proof that the proposed remote illusion device is without any limitation of incident wave's direction. It is important to realize that, since the center region we created is a hidden region, the object to be transformed is not limited to be made of dielectric materials. This shows that the remote illusion device can be applied to any arbitrary object.

In conclusion, we proposed a new recipe to remotely conceal any arbitrary object by applying multi-folded transformation optics. This remote device can hide any object from a certain distance and it can be used to remotely change the field pattern of one object into that of another one. The hidden object can move freely inside the hidden region created by the proposed device, and it can change shape, without requiring the cloak to create different anti-objects for each different structure. Due to the material dispersed, the device we proposed is still limited to a small frequency band. Still, the multi-folded transformation method will be very helpful to design many other remote devices such as remote waveguides, remote antennas, sensors etc., that will be very useful in future microwave and optical applications.

The authors thank P. Rebusco for critical reading and editing of the manuscript. This work was sponsored by the National Natural Science Foundation of China under Grants No. 61322501, No.61574127, and No. 61275183, the Top-Notch Young Talents Program of China, the Program for New Century Excellent Talents (NCET-12-0489) in University, the Fundamental Research Funds for the Central Universities, the Innovation Joint Research Center for Cyber-Physical-Society System and the Postdoctoral Science Foundation of China under Grant No. 2015M581930.

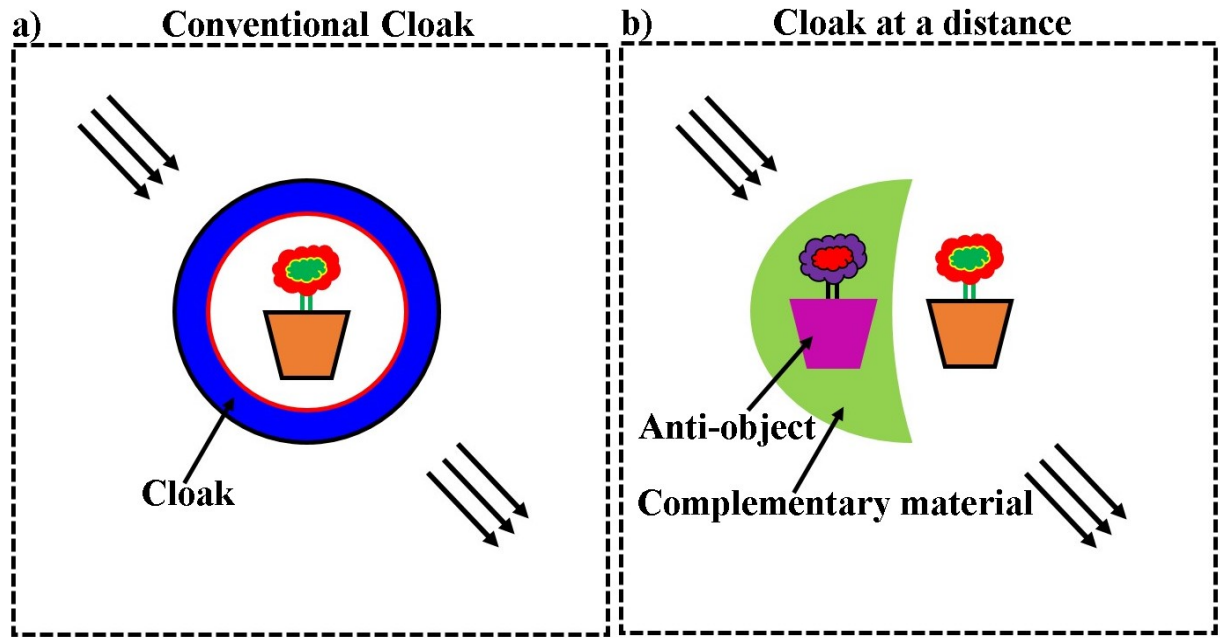

**Figure 1:** Illustration of the ideal conventional cloak and the anti-object remote cloaking device. (a) A conventional cloak [1] is used to imprison the hidden object with its cloaking shell, so that the hidden object is not penetrated by EM waves. (b) An object dependent remote cloaking device proposed by Lai et al. [19] is based on the "anti-object" concept, which is used to conceal the object outside of the cloaking shell, in which a complementary media embedded with a complementary image of the cloaked object is used to cancel the scattering effect of the actual cloaked object.

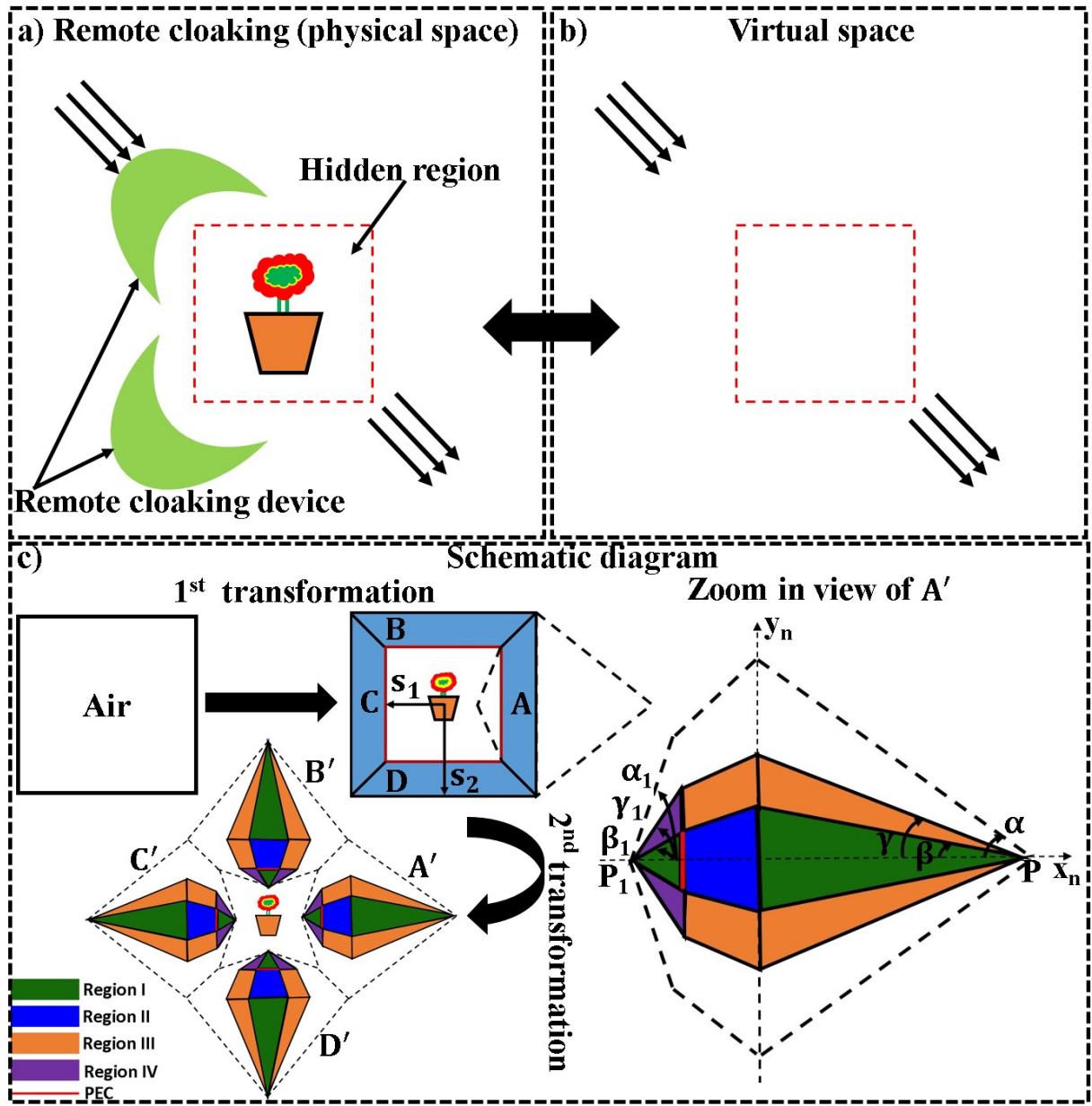

**Figure 2:** Schematic diagram of the proposed object independent remote cloaking device. (a) One or several remote devices can be used to realize the hidden region in the real space. For comparison, the hidden region generated by such device remains similar to that of an ideal cloak (b). (c) Schematic diagram of the proposed remote cloaking device, based on two steps. At the first step the virtual space is transformed into a closed square cloak with four different segments (here labeled $A$, $B$, $C$ and $D$) and then a multi-folded transformation method is applied on each part to make it a remote cloaking device. As an example, after applying the second transformation function, segment $A$ turns into $A'$, which is composed of four different regions. Zoom in view used to clearly visualize the different regions of each part in reference to the derivation in the main text.

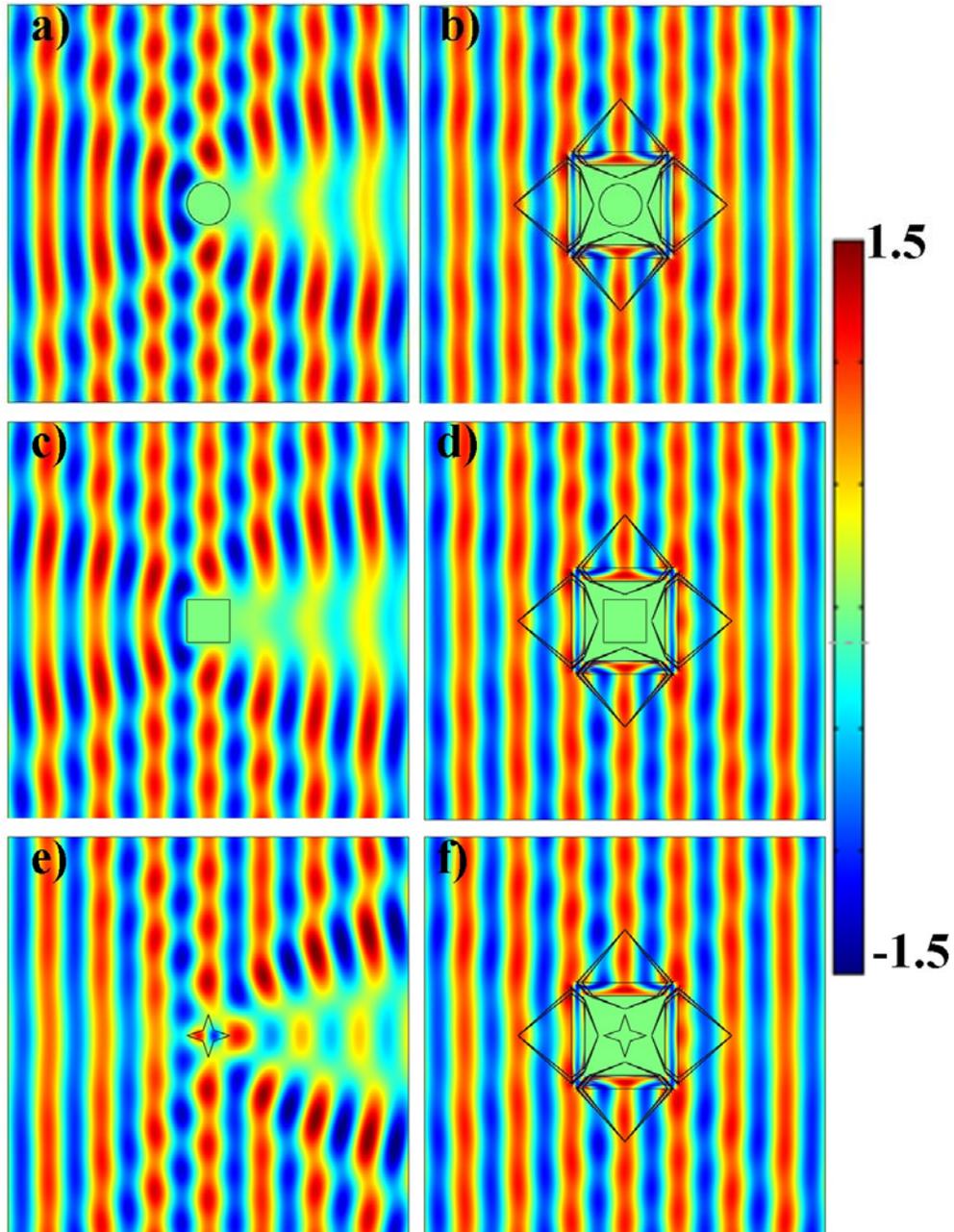

**Figure 3:** Comparison of the total electric field distributions in 2-D TE scattering mode for both an object alone and the same object beside the object independent remote cloaking device. (a) The incoming TE wave is impinging a PEC circular shaped object along the x-direction, producing large scattering, while in (b) the object independent remote cloaking device reduces the scattering effect, achieving to cloak the circle. Similarly, the scattering of a PEC square shaped object in (c) and a dielectric star of $\varepsilon_0 = 5$ and $\mu_0 = 1$ in (e) are reduced in (d) and (f) respectively. This figure validates the object independency concept of the proposed device.

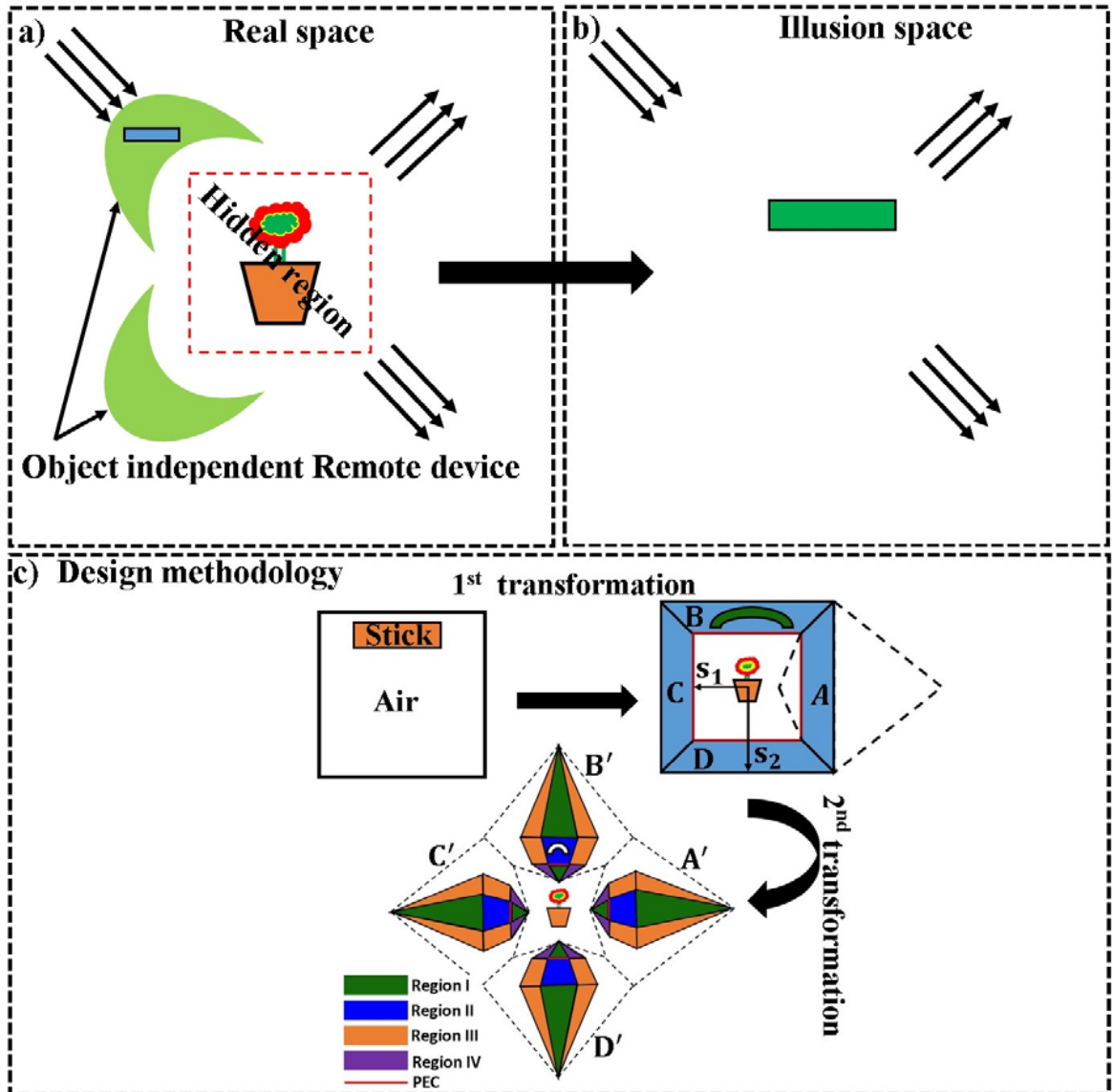

**Figure 4:** Schematics of an object independent illusion device that remotely transforms the image of one object (a flower) into another object (a stick). (a) The flower (any object) is free to move inside the hidden region surrounded by the remote illusion device in real space. (b) The stick (or any object used for illusion) in the illusion space. (c) Designing methodology of the remote illusion device in real space. The remote illusion device is based on the remote cloaking device (Fig. 2) with some minor addition such as a stick (or any object) embedded into the cloak region $B$, originally transformed from virtual space. Next, the stick (or any object) is further compressed to the $B'$ region while maintaining the same domain.

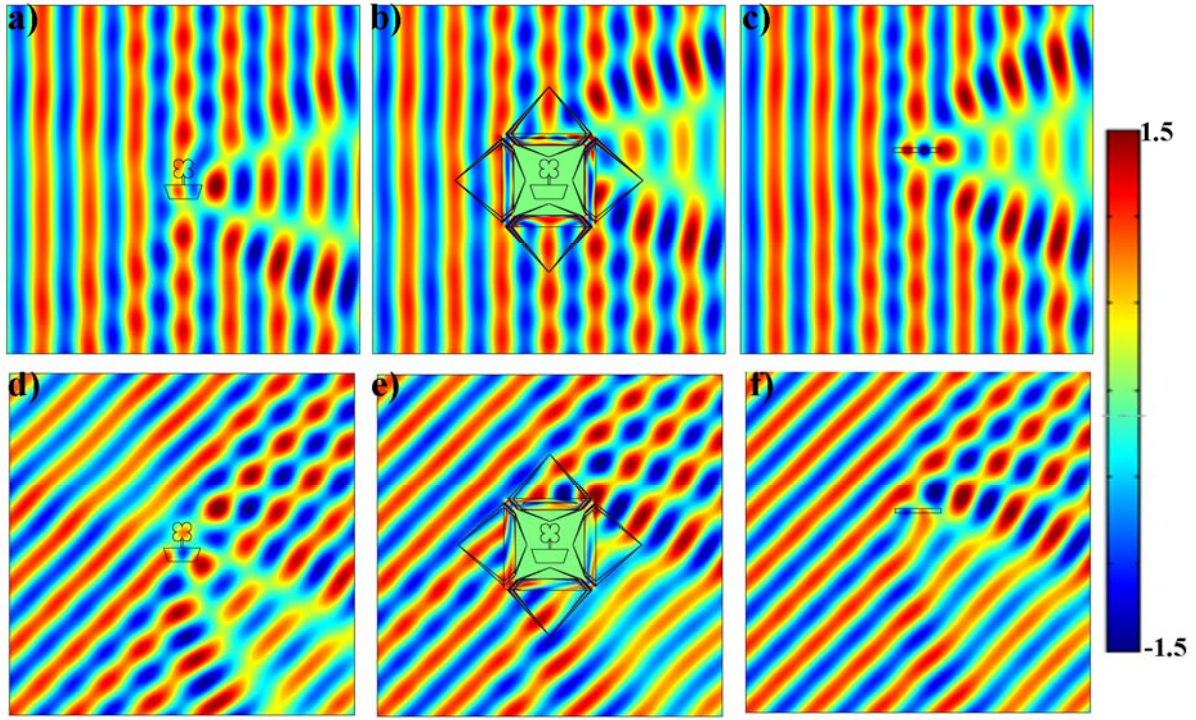

**Figure 5:** The numerical demonstration of remotely transforming the image of a dielectric flower with $\varepsilon_f = 3$ and $\mu_f = 1$ into a stick with $\varepsilon_s = 5$ and $\mu_s = 1$ through an object independent remote illusion device under an incident TE plane wave propagating from left (a-c) and top left (d-f). (a) Scattering behavior of the dielectric flower. (b) Scattering behavior of the dielectric flower besides the object independent remote illusion device, identical to that of the dielectric stick as shown in (c). (d-f) The incident plane wave is rotated by $\pi/4$ with respect to the objects and the remote illusion device. The scattering pattern of (e) is exactly the same as that of (f).